\documentclass{article} 
\usepackage{graphicx}

\begin{document}
\title{The Raychaudhuri Equation for Spinning Test Particles}
\author{Morteza Mohseni\footnote{email: m-mohseni@pnu.ac.ir}\\
              Physics Department, Payame Noor University, 19395-3697 Tehran, Iran}
\maketitle
\begin{abstract}
We obtain generalized Raychaudhuri equations for spinning test particles corresponding to congruences of particle's world-lines,
momentum, and spin. These are physical examples of the Raychaudhuri equation for a non-normalized vector, unit time-like vector, and 
unit space-like vector. We compute and compare the evolution of expansion-like parameters associated with these congruences for spinning particles 
confined in the equatorial plane of the Kerr space-time. 
  
\vspace{5mm}
   
\end{abstract}

\section{Introduction}
The well-known Raychaudhuri equation has played a prominent role in many applications of the general theory of 
relativity. Its application in investigations of space-time singularities is a classic example of this, see \cite{kar,elis} 
for reviews of this equation. A related problem is the study of focusing and defocusing of geodesics, which is of 
particular interest in cosmology \cite{alb}. The interest in the Raychaudhuri equation is not limited to the 
framework of general relativity, and several generalized equations have been obtained in other theories of gravity 
\cite{rip,kar2,shoj,har,grec} and in mathematical physics \cite{val}. Recent applications of the equation include 
those presented in \cite{borg,stav,tsag,alb2}. 

On the other hand, trajectories of particles moving in curved space can deviate from the geodesics of the 
background as a result of non-gravitational interactions with other objects. Then, the study of geodesic focusing gets 
replaced by the study of focusing or defocusing of world-lines and the Raychaudhuri equation gets modifications 
accordingly. This idea has been used in several situations, e.g.,  in \cite{tsagas} where the gravitational collapse of a 
magnetized medium is studied, and in \cite{kou} where the same problem was considered for a charged medium.  

An interesting example of non-geodesic motion is the motion of  particles with internal structures. In the simplest 
situation, the case of spinning particles, where the particle has an internal angular momentum, the particle path is 
influenced by the coupling between its spin and the background curvature. In the so-called pole-dipole 
approximation, this is described by the Mathisson-Papapetrou-Dixon (MPD) equations \cite{dix}, which we will 
briefly review in the next section. These equations have been used to study the trajectories of spinning particles in 
different space-time backgrounds \cite{hle1,yuan,ali,bini2,gera,ply,ni1,bini,hor,sem2,plat}. The equation is also of 
interest within the context of extended theories of gravity \cite{mohprd,put,obu}. A generalization of the geodesic 
deviation equation for spinning particles was presented in \cite{nieto,mohplb}.

The idea of including spin in the Raychaudhuri equation has been considered in \cite{smal} for a spinning fluid. 
Here, we are interested in obtaining extended versions of the Raychaudhuri equation for spinning particles. Such 
generalized equations would be of importance in investigating the focusing of spinning particle world-lines. The 
standard technique of obtaining the Raychaudhuri equation in which the expansion, the shear, and the vorticity 
tensors are constructed out of a normalized time-like four-velocity (for massive particles), does not work here, 
because the MPD equations do not guarantee the existence of such a four-vector \cite{ehl}. Here, we utilize the 
method developed recently in \cite{abre}, by which the above-mentioned construction may be made for a 
non-normalized vector.

In the following sections, we first review the equations of motion of spinning particles in the so-called pole-dipole approximation.
We then obtain versions of the Raychaudhuri equation associated with different physically meaningful vectors relevant to spinning 
particles, i.e., the spin, the momentum, and the velocity vectors. Then we investigate the applications of these generalized Raychaudhuri 
equations by considering a congruence of spinning particle world-lines confined to the equatorial plane of the Kerr space-time which 
provides a rather rare exmaple of exact solutions to the equations of motion. We obtain the explicit expressions for the expansion scalar 
associated with spin, momentum, and velocity vectors. In the last section, we discuss the results. 
\section{The Equations of Motion}
By setting $s^{\mu\nu}, p^\mu,$ and $v^\mu$, to represent the particle spin tensor, momentum four-vector, and 
velocity respectively, one can write the MPD equations in the following form \cite{dix}
\begin{eqnarray}
{\dot s^{\mu\nu}}&=&p^\mu v^\nu-p^\nu v^\mu,\label{eq3a}\\
{\dot p^\mu}&=&-\frac{1}{2}{R^\mu}_{\nu\alpha\beta}v^\nu s^{\alpha\beta},\label{eq3b}
\end{eqnarray}
In the above set of equations, $R_{\mu\nu\kappa\lambda}$ is the curvature tensor, and over-dots stand for 
covariant differentiation $v^\alpha\nabla_\alpha$. To 
complete the equations of motion; we supplement them by the so-called Tulczyjew condition
\begin{equation}\label{eq4}
p_\mu s^{\mu\nu}=0.
\end{equation}
Under these equations, the mass and spin of the particle are conserved, i.e.,
\begin{eqnarray}
p_\mu p^\mu&=&\mbox{const.}=-m^2,\label{h1}\\
s_{\mu\nu}s^{\mu\nu}&=&\mbox{const.}=2s^2.\label{h2}
\end{eqnarray}
The above equations result in the following relation for the velocity \cite{ehl}
\begin{equation}\label{eq11}
v^\mu=\frac{v_\kappa p^\kappa}{p_\lambda p^\lambda}\left(p^\mu+\frac{2s^{\mu\nu}R_{\nu\rho\alpha\beta}
s^{\alpha\beta}p^\rho}{-4p_\gamma 
p^\gamma+s^{\delta\eta}R_{\delta\eta\epsilon\zeta}s^{\epsilon\zeta}}\right)
\end{equation}
which can be simplified by choosing specific gauges, say,  
\begin{equation}\label{tet4}
v_\kappa p^\kappa=-m
\end{equation}
for which the instantaneous zero-momentum and the zero-velocity frames are simultaneous \cite{dix}. By defining $\frac{1}{m}p^\mu=\pi^\mu$, one 
can show that $f^\mu=v^\mu-\pi^\mu$ is orthogonal to $\pi^\mu$ and can be interpreted as the particle 3-velocity with respect to the zero-momentum observer 
whose 4-velocity is $\pi^\mu$,  \cite{ehl}. According to Eq. (\ref{eq11}), the velocity squared $v^\mu v_\mu$  is not constant, and in fact, it can be space-like or 
null in some situations \cite{ehl}.  
In the above gauge, we rewrite Eq. (\ref{eq11}) in the following form 
\begin{equation}\label{e5}
v^\mu=\pi^\mu+f^\mu
\end{equation}
with
\begin{equation}\label{e5a}
f^\mu=\frac{2s^{\mu\nu}R_{\nu\rho\alpha\beta}
s^{\alpha\beta}}{4m^2+s^{\delta\eta}R_{\delta\eta\epsilon\zeta}s^{\epsilon\zeta}}\pi^\rho
\end{equation}
It is also convenient to represent the particle spin with the spin four-vector $s^\mu$ defined by
\begin{equation}\label{m1}
s^\mu=\frac{1}{2\sqrt{-g}}\epsilon^{\mu\nu\alpha\beta}\pi_\nu s_{\alpha\beta}
\end{equation}
in which $\epsilon^{\mu\nu\alpha\beta}$ is the alternating symbol.
\section{The generalized Raychaudhuri equations}
For a congruence of time-like world-lines with unit tangent $u^\mu=\frac{dx^\mu}{d\tau}$, the Raychaudhuri 
equation is given by 
\begin{equation}\label{m2}
{\dot\Theta}=-R_{\mu\nu}u^\mu u^\nu+\omega_{\mu\nu}\omega^{\mu\nu}-\sigma_{\mu\nu}\sigma^{\mu\nu}
-\frac{1}{3}\Theta^2+\nabla_\mu{\dot u^\mu}
\end{equation}
where an over-dot means $\frac{D}{D\tau}=u^\mu\nabla_\mu$ ($\nabla_\mu$ being the covariant derivative), 
$$\theta^{\mu\nu}=\frac{1}{2}h^{\mu\kappa}h^{\lambda\nu}
(\nabla_\kappa u_\lambda+\nabla_\lambda u_\kappa)$$ is the expansion tensor, 
$\Theta=g_{\mu\nu}\theta^{\mu\nu}$, 
$$\sigma_{\mu\nu}=\theta_{\mu\nu}-\frac{1}{3}h_{\mu\nu}\Theta$$ is the shear tensor, 
$$\omega^{\mu\nu}=\frac{1}{2}h^{\mu\kappa}h^{\lambda\nu}
(\nabla_\kappa u_\lambda-\nabla_\lambda u_\kappa)$$ is the 
vorticity (rotation) tensor, and 
\begin{equation}\label{m2a}
h_{\mu\nu}=g_{\mu\nu}+u_\mu u_\nu.
\end{equation}
When the congruence is geodesic, the last term in the right-hand side of the above equation disappears, and one is 
left with the standard text-book form of the equation. 

Now, considering a congruence of spinning particle world-lines, the above equation may be generalized in different 
ways by taking various relevant objects $\pi^\mu, v^\mu$, and $s^\mu$ into 
account. The simplest situation we face with, is the one in which $u^\mu$ in (\ref{m2}) is replaced by $\pi^\mu$. This results in
\begin{eqnarray}
\pi^\lambda\nabla_\lambda\Theta=-R_{\mu\nu}\pi^\mu\pi^\nu+\omega_{\mu\nu}\omega^{\mu\nu}-\sigma_{\mu\nu}\sigma^{\mu\nu}-\frac{1}
{3}\Theta^2+\nabla_\mu({\pi^\lambda\nabla_\lambda\pi^\mu}).\label{eq11a}
\end{eqnarray}
This is the Raychaudhuri equation for a time-like congruence corresponding to the world-lines of  zero-momentum observers with 4-velocity 
$\pi^\mu$. By using Eqs. (\ref{e5}) and (\ref{eq3b}), this can also be rewritten in the following form  
\begin{eqnarray}\label{x1}
v^\lambda\nabla_\lambda\Theta&=&-R_{\mu\nu}v^\mu\pi^\nu+\omega_{\mu\nu}\omega^{\mu\nu}-\sigma_{\mu\nu}\sigma^{\mu\nu}-\frac{1}
{3}\Theta^2-\nabla_\mu f^\lambda\nabla_\lambda\pi^\mu\nonumber\\&&-\nabla_\mu\left(\frac{1}{2m}{R^\mu}_{\nu\alpha\beta}v^\nu s^{\alpha\beta}\right)
\end{eqnarray}
which can be used to calculate the rate of change of $\Theta=\nabla_\alpha\pi^\alpha$ along $v^\mu$. The last term in this equation shows the explicit dependence
on spin. There are also implicit spin dependences through $v^\mu$ and $f^\mu$. Expressed in terms of the normalized time-like vector $\pi^\mu$ and the non-normalized 
vector $v^\mu$, this may also be considered as a physical example of the equation constructed in section VI of Ref. \cite{abre}.

Regarding the velocity vector field $v^\mu$, which is non-normalized here, one cannot use the standard Raychaudhuri equation. However, it can be described via a 
generalized Raychaudhuri equation introduced in Ref. (\cite{abre}). In a slightly recast shape, it can be written in the following form
\begin{eqnarray}\label{abr1} 
v^\lambda\nabla_\lambda(\nabla_\kappa v^\kappa)&=&-R_{\alpha\beta}v^\alpha v^\beta-\frac{2}{3}(\nabla_\alpha v^\alpha)^2+\nabla_\alpha(v^\lambda\nabla_\lambda v^\alpha)
\nonumber\\&&+\frac{1}{4}(\nabla_\alpha v_\beta+\nabla_\beta v_\alpha)(\nabla^\alpha v^\beta+\nabla^\beta v^\alpha).
\end{eqnarray}
It is also possible to establish a useful relationship between $\nabla_\mu\pi^\mu$ and $\nabla_\mu v^\mu$ by using Eq. (\ref{e5}). It reads
\begin{equation}\label{tt1}
\nabla_\mu v^\mu = \nabla_\mu\pi^\mu+\nabla_\mu(F^\mu_\rho \pi^\rho)
\end{equation} 
where $F^\mu_\rho=\frac{2s^{\mu\nu}R_{\nu\rho\alpha\beta}
s^{\alpha\beta}}{4m^2+s^{\delta\eta}R_{\delta\eta\epsilon\zeta}s^{\epsilon\zeta}}$.

On the other hand, getting back to Eq. (\ref{eq11}), one can argue that those points in which $v^\mu$ becomes null or space-like correspond to situations where 
pole-dipole approximation (upon which the equations of motion are based, see e.g. Ref. \cite{papa}) is not valid. Thus, considering the limits of validity of the 
pole-dipole approximation, one can define the unit time-like vector 
\begin{equation}
u^\mu=\frac{v^\mu}{\sqrt{1-f_\nu f^\nu}},\label{d1}
\end{equation}
in which the condition $f_\mu f^\mu<1$ is assumed. By inserting this back into Eq. (\ref{m2}), another version of  the Raychaudhuri equation may be obtained.

Still, another version of the Raychaudhuri equation may be obtained by taking the unit space-like vector $\sigma^\mu\equiv\frac{s^\mu}{s}$ into account. 
This can be obtained by using $h_{\mu\nu}=g_{\mu\nu}-\sigma_\mu \sigma_\nu$ instead of the projection tensor defined by Eq. (\ref{m2a}). Inserting this and 
replacing $u^\mu$ by $\sigma^\mu$ in Eq. (\ref{eq11}) then results in the desired equation. However, as discussed in Ref. (\cite{abre}), this would be a formal 
generalization without the usual meaning of the Raychaudhuri equation. An interesting point is that $\sigma^\mu$ is orthogonal to the time-like hypersurfaces
spanned by $u^\mu$ and the resulting equation may be re-expressed in terms the extrinsic curvature of these hypersurfaces via the technique introduced in Ref. 
\cite{abre}.  
\section{Applications}
One of the main applications of  the Raychaudhuri equation is in the study of space-time singularities. A powerful tool in this regard is an 
offspring of this equation, the so-called focusing theorem. In its simplest form, it sates that in a vorticity-free space-time (i.e. $\omega_{\mu\nu}=0$), the 
geodesics focus provided the 
null and strong energy conditions hold \cite{hawk}. This is easily seen from the following equation
\begin{equation}
{\dot\Theta}=-R_{\mu\nu}u^\mu u^\nu-\sigma_{\mu\nu}\sigma^{\mu\nu}-\frac{1}{3}\Theta^2\label{d4}
\end{equation}
in which the right-hand side is negative. 

In the same manner, one can obtain a similar relation for focusing of spinning particles by dropping $\omega^2$
in Eq. (\ref{x1}). However, due to the presence of spin dependent terms, there is no guarantee that the right-hand side of the resulted equation is always negative. 
To proceed further, here we consider a solution of the equations of motion (\ref{eq3a})-(\ref{eq4}) which was introduced in Ref. \cite{tod} and is among 
the very few known exact solutions. It describes the motion of spinning particles in the equatorial plane of the Kerr black hole\footnote{By the way, it seems 
that there is an apparent mix-up of coordinate systems in \cite{tod} in which the metric was expressed in terms of the Boyer-Lindquist system while the Kinnersley 
null-tetrad was used.}. In terms of the coordinates used in Ref. \cite{kin}, the metric is given by  
\begin{eqnarray}
ds^2&=&-\left(1-\frac{2Mr}{\Sigma}\right)dt^2-\frac{4Mar\sin^2\theta}{\Sigma}dtd\phi-2dtdr+\Sigma d\theta^2
\nonumber\\&&+2a\sin^2\theta drd\phi+\frac{(a^2+r^2)^2-a^2\sin^2\theta\Delta}{\Sigma}\sin^2\theta d\phi^2\label{d4a}.
\end{eqnarray}
where $\Sigma=r^2+a^2\cos^2\theta$, $\Delta=r^2+a^2-2Mr$, and $M, a$ are the mass and the specific angular momentum of the source. The explicit form of 
the solution is given in terms of the null tetrad $(l^\mu,n^\mu,m^\mu,{\bar m}^\mu)$ (due to Kinnersley \cite{kin}) by
\begin{eqnarray}
p^\mu&=&q l^\mu+b n^\mu+cq^\mu,\label{d3b}\\
s^\mu&=&\frac{s}{\sqrt 2}(m^\mu+{\bar m}^\mu)\label{d4b}\\
v^\mu&=&A l^\mu+B n^\mu+Cq^\mu,\label{d3c}
\end{eqnarray}
where
\begin{eqnarray}
l^\mu&=&(0,1,0,0),\label{tet1}\\
n^\mu&=&\frac{1}{\Sigma}(r^2+a^2,-\frac{1}{2}\Delta,0,a),\label{tet2}\\
m^\mu&=&\frac{-\bar\rho}{{\sqrt 2}}\left(ia\sin\theta,0,1,\frac{i}{\sin\theta}\right)\label{tet3}
\end{eqnarray}
with $\rho=-\frac{1}{r-ia\cos\theta}$, and $q^\mu =\frac{i}{\sqrt 2}(m^\mu-{\bar m}^\mu)$. The constant $q, b, c, A, B, C$
should satisfy $2qb-c^2=m^2$, and $Ab+Bq-cC=m$. For $\theta=\frac{\pi}{2}$, the momentum $p^\mu $ and the velocity 
$v^\mu$ are confined to the equatorial plane and $s^\mu$ is perpendicular to it. Using Eq. (\ref{eq11}), we can obtain the following expressions for
the components of $v^\mu$ for $\theta=\frac{\pi}{2}$  
\begin{eqnarray*}
A&=&\left(1+\frac{3Ms^2(2bq-m^2)}{m^4r^3-2Ms^2(3bq-m^2)}\right)\frac{q}{m}\\
B&=&\left(1+\frac{3Ms^2(2bq-m^2)}{m^4r^3-2Ms^2(3bq-m^2)}\right)\frac{b}{m}\\
C&=&\left(1+\frac{6Ms^2qb}{m^4r^3-2Ms^2(3bq-m^2)}\right)\frac{\sqrt{2qb-m^2}}{m}.
\end{eqnarray*}
There is also a solution with $\sqrt{2bq-m^2}\rightarrow -\sqrt{2bq-m^2}$. It is interesting to note that $A$, $B$, and $C$ above are independent of $a$.

Now, we first insert the unit space-like vector $$\sigma^\mu=\frac{1}{s}s^\mu=\frac{1}{\sqrt 2}(m^\mu+{\bar m}^\mu)$$ into Eq. (\ref{m2}). For 
$\theta=\frac{\pi}{2}$, we obtain 
$$R_{\alpha\beta}\sigma^\alpha\sigma^\beta=\sigma_{\alpha\beta}\sigma^{\alpha\beta}=\omega_{\alpha\beta}\omega^{\alpha\beta}=\Theta_{\alpha\beta}
\Theta^{\alpha\beta}=0,$$ and the equation reduces to
\begin{eqnarray*}
\sigma^\alpha\nabla_\alpha\Theta=-\frac{1}{r^2}=\nabla_\beta(\sigma^\alpha\nabla_\alpha\sigma^\beta)
\end{eqnarray*}
and the generalized Raychaudhuri equation is reduced to an equation for the acceleration-like quantity $\sigma^\alpha\nabla_\alpha\sigma^\beta$
in this case.

For unit time-like vector $\pi^\mu=\frac{1}{m}(ql^\mu+bn^\mu+{\sqrt{2qb-m^2}}q^\mu)$, 
one can insert the relevant quantities in Eq. (\ref{eq11a}) to show that the Raychaudhuri equation is satisfied. In particular, we have 
\begin{equation}\label{f1}
\Theta=\frac{Mb+(2q-b)r-a{\sqrt{2bq-m^2}}}{mr^2}
\end{equation}
and
\begin{equation}\label{o1}
\pi^\mu \nabla_\mu \Theta=\frac{1}{2m^2r^5}(a^2b-2Mbr-(2q-b)r^2)(2bM-2a{\sqrt{2qb-m^2}}+(2q-b)r).
\end{equation}
For the special case where $2q=b$, this gets simplified and we obtain the LHS as $\frac{b^2}{m^2r^5}(a^2-2Mr)(M-am{\sqrt{\delta^2-1}})$ in which 
$\delta=\frac{b}{m}$. Now, the expression inside the first parenthesis is negative and hence the congruence is convergent whenever $M$ is larger than 
$am{\sqrt{\delta^2-1}}$.

\begin{figure}[h]
\begin{center}
\includegraphics[width=5.8cm]{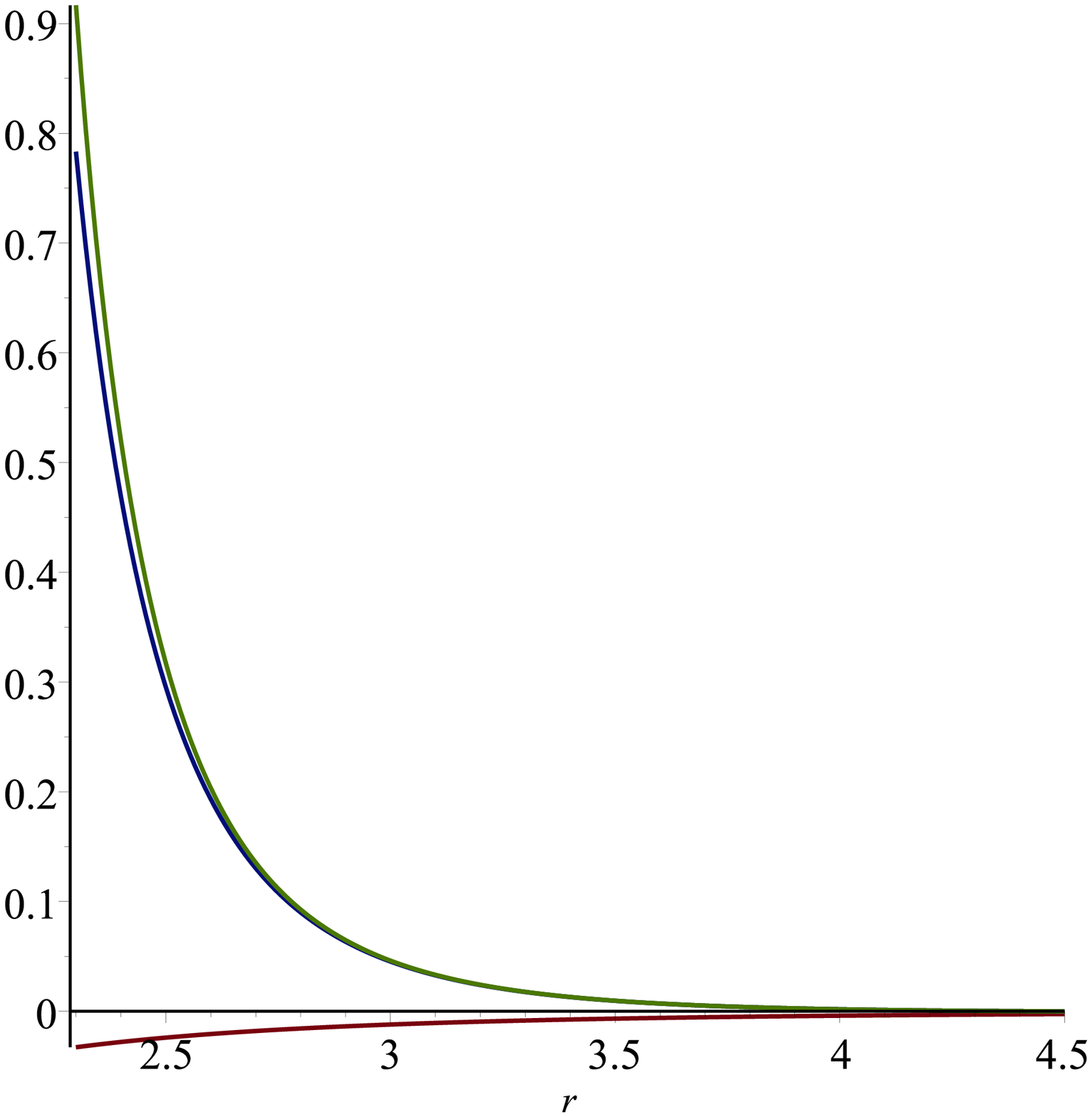}\includegraphics[width=5.8cm]{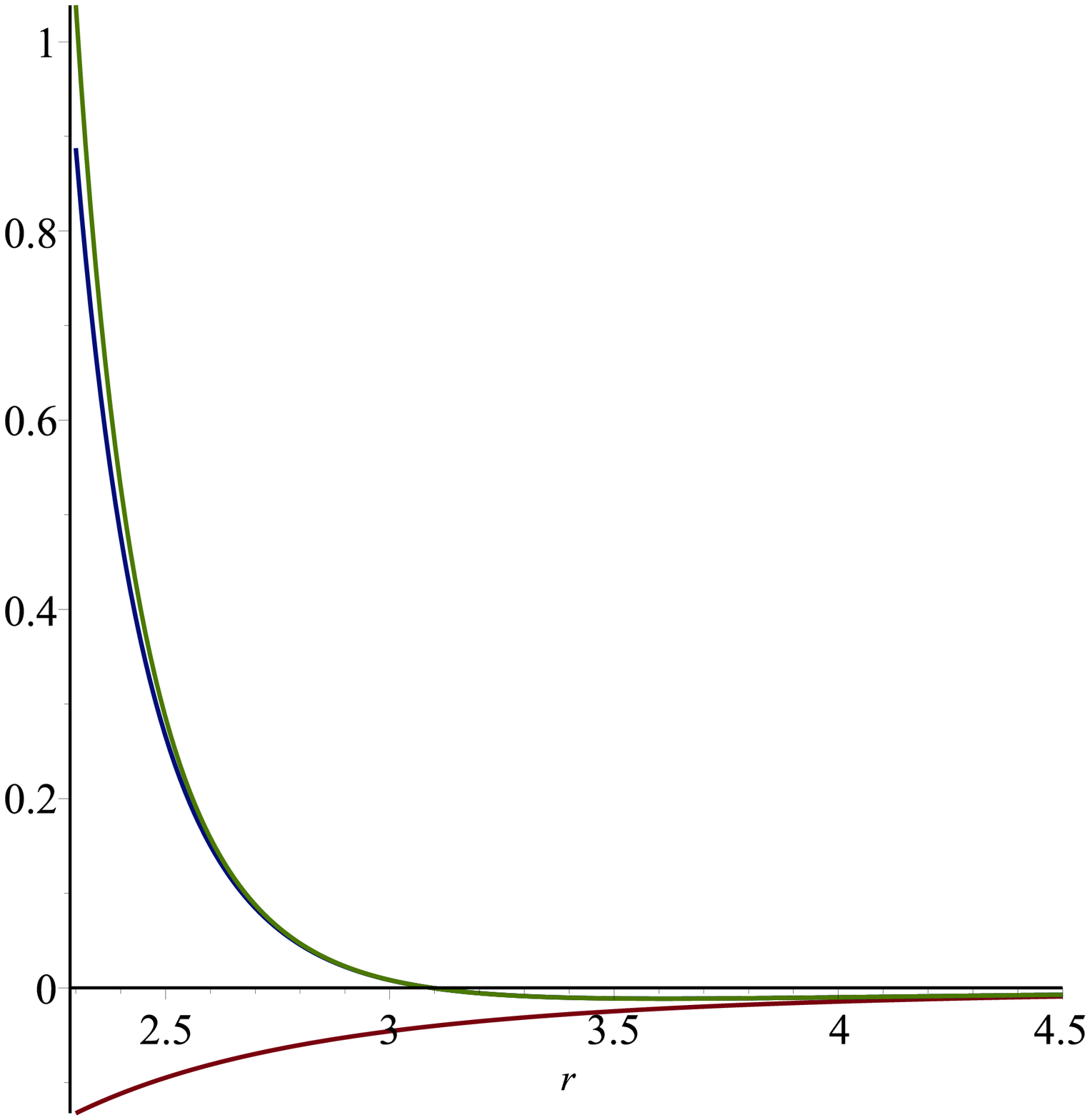}
\end{center}
\caption{The quantities $\pi^\lambda\nabla_\lambda(\nabla_\mu\pi^\mu)$ (lower),  $v^\lambda\nabla_\lambda(\nabla_\mu v^\mu)$ (middle), and  
$u^\lambda\nabla_\lambda(\nabla_\mu u^\mu)$ (upper) in terms of $r$ for $\frac{s}{m}=1, q=\frac{m}{2}, b=m, M=1$, $a=1$ (left), and $a=0.1$ (right).}\label{fig1} 
\end{figure}

\begin{figure}
\begin{center}
\includegraphics[width=5.8cm]{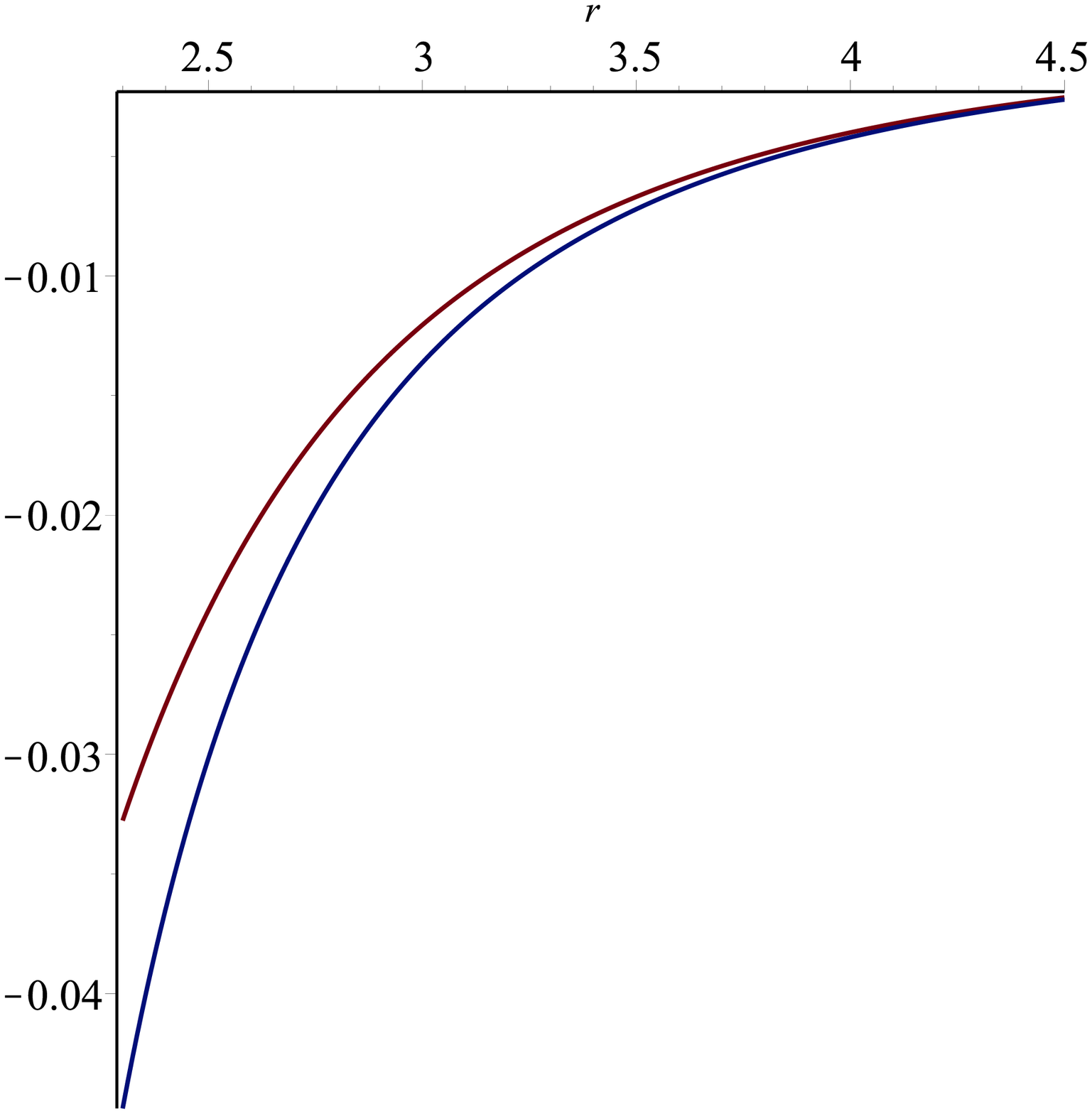}\includegraphics[width=5.8cm]{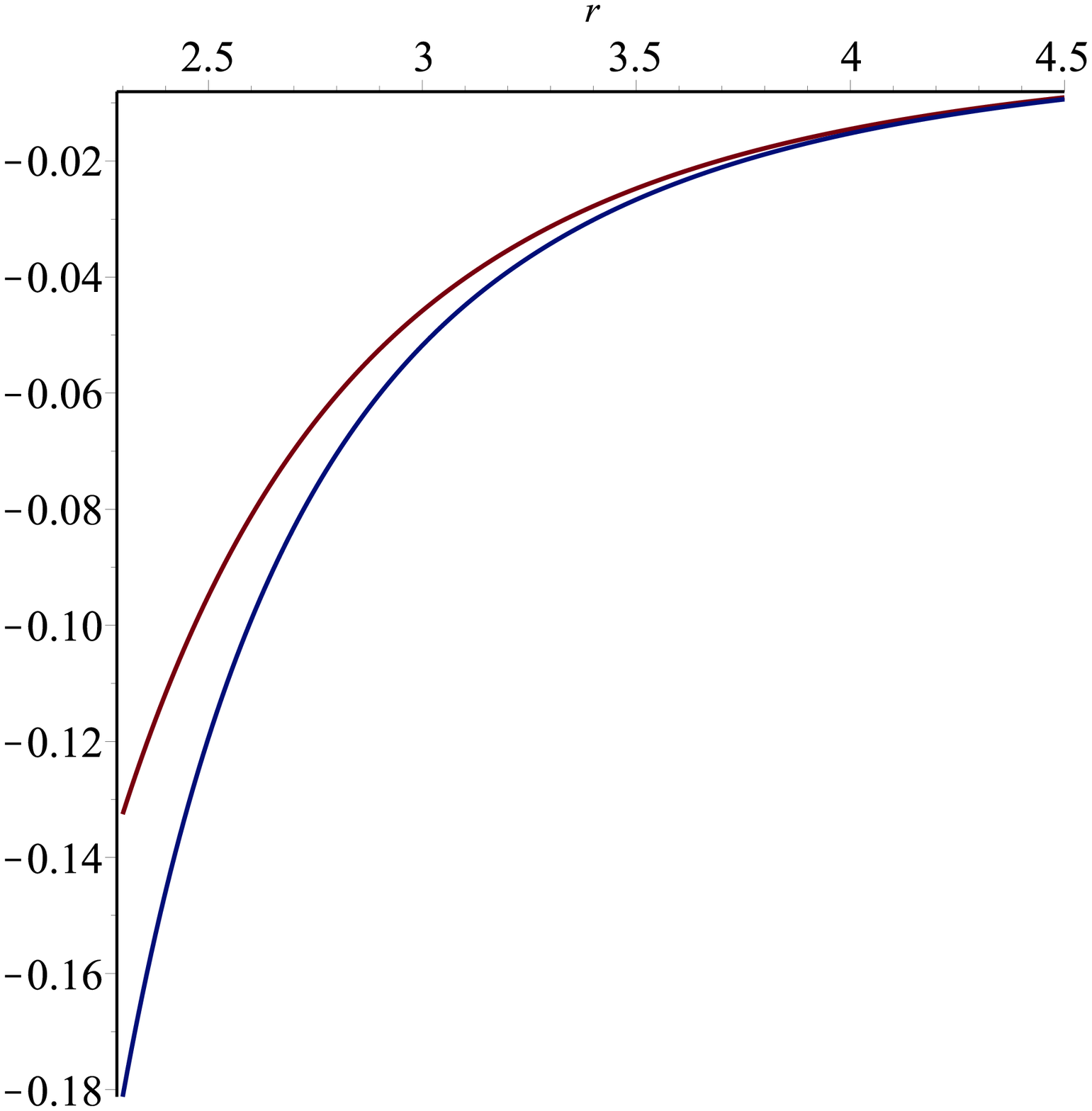}
\end{center}
\caption{The quantities $\pi^\lambda\nabla_\lambda(\nabla_\mu\pi^\mu)$ (upper) and $v^\lambda\nabla_\lambda(\nabla_\mu\pi^\mu)$ (lower) in terms of $r$ for $\frac{s}{m}=1, 
q=\frac{m}{2}, b=m, M=1$, $a=1$ (left), and $a=0.1$ (right).}\label{fig2} 
\end{figure}

We can also insert Eq. (\ref{d3c}) into Eq. (\ref{abr1}) to check that it holds.  Finally, one can examine
the Raychaudhuri equation for $u^\mu=\frac{1}{\sqrt{-v_\nu v^\nu}}v^\mu$. The resulting expressions are algebraically complicated.
The behavior of $\pi^\lambda\nabla_\lambda(\nabla_\mu\pi^\mu)$, $v^\lambda\nabla_\lambda(\nabla_\mu v^\mu)$, $u^\lambda\nabla_\lambda(\nabla_\mu u^\mu)$, and
$v^\lambda\nabla_\lambda(\nabla_\mu\pi^\mu)$ are sketched in Figs. \ref{fig1} and \ref{fig2}. 

The figures show that the congruence described by the velocity vectors $v^\mu$ or $u^\mu$ diverges for extreme Kerr solution while the congruence corresponding to 
the momentum vector converges. For the limit of small values of $a$ both congruences converge above certain values of $r$. 

\begin{figure}
\begin{center}
\includegraphics[width=5.8cm]{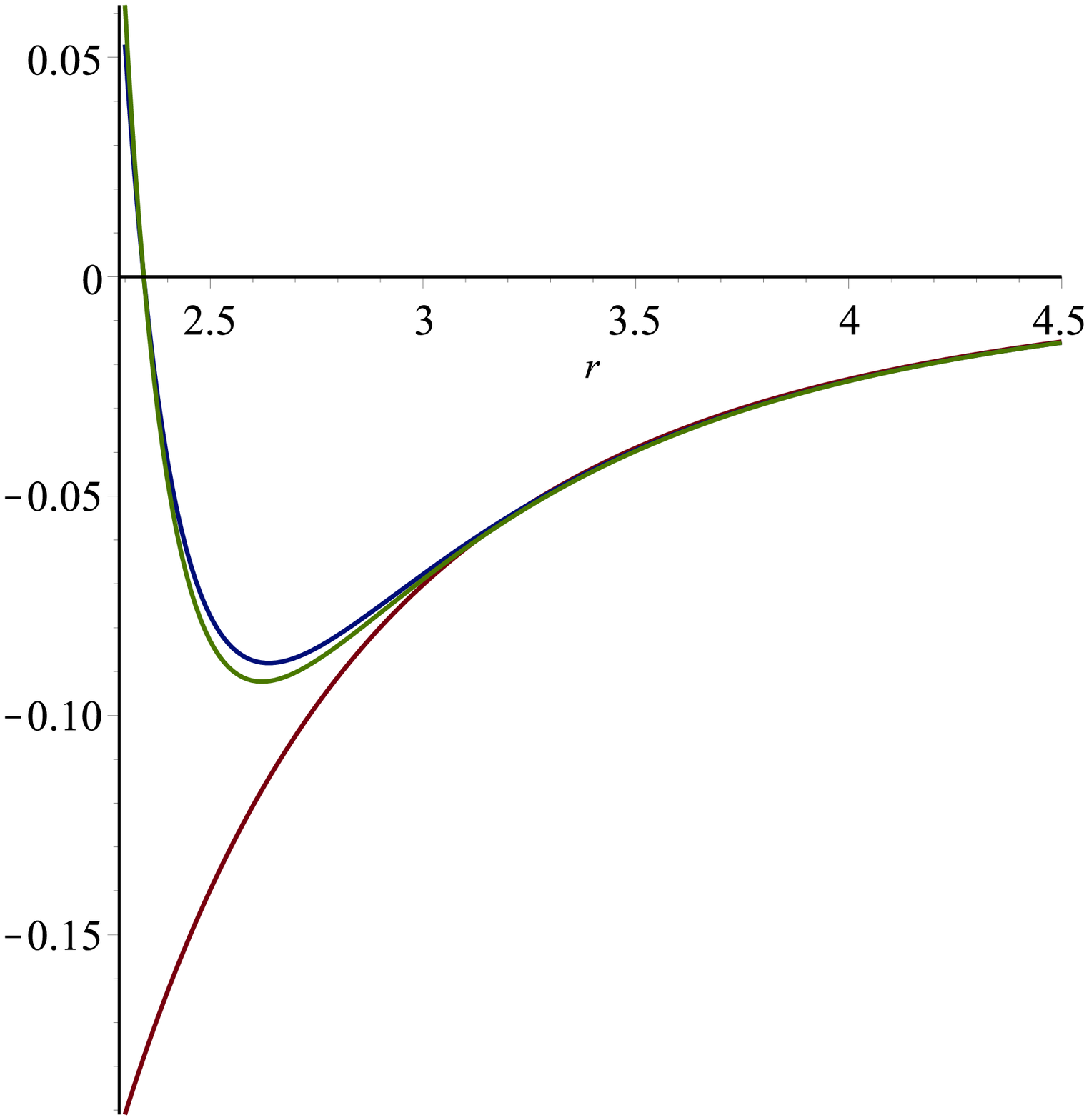}\includegraphics[width=5.8cm]{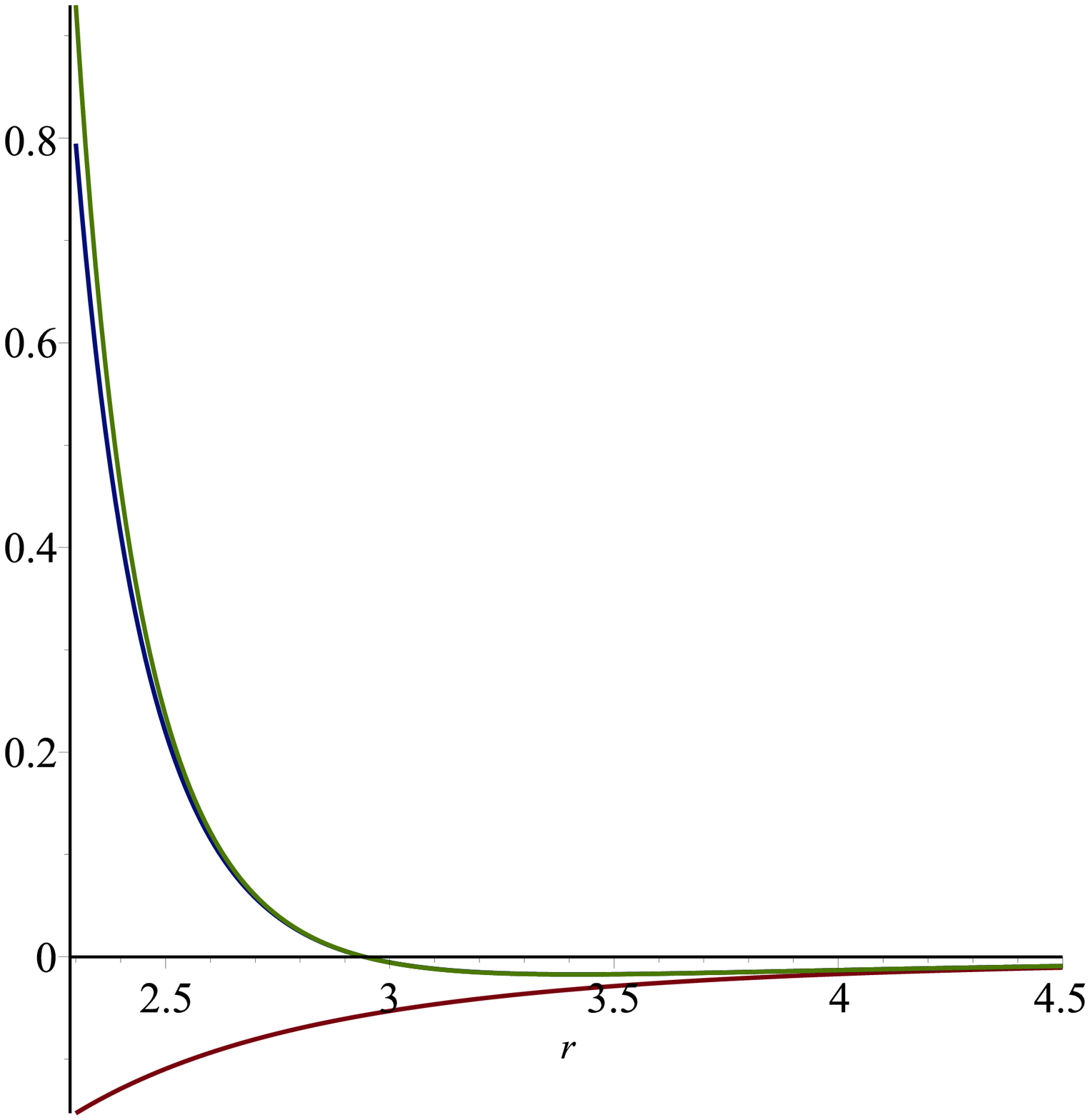}
\end{center}
\caption{The quantities ${\bar\pi}^\lambda\nabla_\lambda(\nabla_\mu{\bar\pi}^\mu)$ (lower),  ${\bar v}^\lambda\nabla_\lambda(\nabla_\mu {\bar v}^\mu)$ (upper), and  
${\bar u}^\lambda\nabla_\lambda(\nabla_\mu {\bar u}^\mu)$ (middle) in terms of $r$ for $\frac{s}{m}=1, q=\frac{m}{2}, b=m, M=1$, $a=1$ (left), and $a=0.1$ (right)
.}\label{fig3} 
\end{figure}

The solution 
\begin{equation}
{\bar\pi}^\mu=\frac{1}{m}(ql^\mu+bn^\mu-{\sqrt{2bq-m^2}}q^\mu)
\end{equation}
results in rather drastically different behavior for the extreme case, which is shown in Fig. \ref{fig3}.
\section{Discussion}
We have obtained generalized versions of the Raychaudhuri equation for a congruence of spinning test particle world-lines. Such a system provides at the same time
normalized space-like and time-like vectors corresponding to the spin and momentum vector fields, and non-normalized vector corresponding to the velocity. At the 
formal level, this can be considered as a physical manifestation of the formalism introduced recently in the literature \cite{abre}.
From the other side, i.e., the spinning particle dynamics, these extended equations might be helpful in a better understanding of the motion of spinning particles.
Even though the complicated behavior of some terms in the right-hand sides of these generalized equations prevents us to deduce general focusing theorems, they still
might be used in particular situations to study the focusing or de-focusing of world-lines.

We have also considered the motion of spinning particles in the equatorial plane of the Kerr space-time by computing the evolution of the expansion-like parameters 
associated with different congruences mentioned above. The existence of several congruences expanding in different manners in a single physical system is an 
intersting phenomena by itself. The above equations may also be obtained for spinning particles described by other known supplementary conditions.

\section*{Acknowledgments}
I would like to thank Payame Noor University for financial support.

\end{document}